\documentclass[aps,prl,twocolumn,showpacs,superscriptaddress]{revtex4}
\usepackage{bm,amsmath,amssymb,latexsym,mathrsfs,enumerate}
\usepackage{graphicx}
\renewcommand{\vec}[1]{\bm{#1}}

\begin{document}

\title{Pairing of Solitons in Two-Dimensional \boldmath $S=1$  Magnets}

\author{B. A. Ivanov}
\affiliation{Institute of Magnetism, National Academy of Sciences and Ministry
  of Education,  03142 Kiev, Ukraine} 
\affiliation{T. Shevchenko Kiev National University,  03127
  Kiev, Ukraine}

\author{R. S. Khymyn}
\affiliation{T. Shevchenko Kiev National University,  03127 Kiev,
Ukraine}

\author{A. K. Kolezhuk}
\thanks{On leave from: Institute of Magnetism, Natl. Acad. Sci. and Ministry
  of Education,  03142 Kiev, Ukraine}
\affiliation{Institut f\"ur Theoretische Physik C, RWTH Aachen,
D-52056 Aachen, Germany}

\date\today

\begin{abstract}
We discuss the structure of topological solitons in a general non-Heisenberg model of isotropic two-dimensional
magnet with spin $S=1$, in the vicinity of a special point where the model
symmetry is enhanced to $SU(3)$.  It is shown that upon perturbing the $SU(3)$
symmetry, solitons with odd topological charge become unstable and bind into
pairs.
\end{abstract}


\maketitle

\emph{Introduction.--} Many condensed matter systems can be
successfully described with the help of effective continuum field
models. In systems with reduced spatial dimensionality,
topologically nontrivial field configurations are known to play an
important role \cite{manton}. Magnetic systems are usually modeled
with the help of the Heisenberg exchange interaction. In many instances the
fluctuations of the length of the local magnetic moment occur at a large energy
scale and can be neglected; the 
continuum field theory in that case is the so-called $O(3)$
nonlinear sigma model (NLSM) describing the dynamics of a
three-component real unit vector field, and the topological
excitations of this model are well understood \cite{Kosevich+All}.

However, for spin $S>1/2$ the general isotropic exchange goes beyond the purely
Heisenberg interaction bilinear in spin operators $\vec{S}_{i}$, and may include
higher-order terms of the type $( \vec{S}_i \vec{S}_j)^n$ with $n$ up to
$2S$. Particularly, a general $S=1$ model with the isotropic 
nearest-neighbor exchange on a
two-dimensional (2d) square lattice is described by the Hamiltonian
\begin{equation}
\label{ham}
\mathcal{H}=-\textstyle\sum_{\langle ij \rangle} h_{ij}, \quad
h_{ij}= {J\left( {{\rm {\bf S}}_i {\rm {\bf S}}_j } \right) +
K\left( {{\rm {\bf S}}_i {\rm {\bf S}}_j } \right)^2},
\end{equation}
where $\langle ij\rangle$ denotes the sum over nearest neighbors, and $J$ and
$K>0$ are respectively the bilinear (Heisenberg) and biquadratic exchange
constants.  The model (\ref{ham}) has been discussed recently in connection with
$S=1$ bosonic gases in optical lattices \cite{ImambekovLukinDemler03} and in the
context of the deconfined quantum criticality
\cite{HaradaKawashimaTroyer07,GroverSenthil07}.  The effective field theory for
the above model is generally more complicated than NLSM: the order parameter
belongs to the 2d complex projective space $CP^{2}$, and at two points, $J=K$
and $J=0$, the model symmetry is enlarged to $SU(3)$.  The aim of the present
paper is to show that the crossover from $SU(3)$ to $SU(2)$ symmetry, taking
place in the vicinity of those special points, features drastic changes in the
structure of topological excitations, which correspond to pairing of solitons of
the $CP^{2}$ model.

\emph{Continuum field description.--} The spin-1 state
$|\psi\rangle_{j}$ at a given site $j$ is a linear superposition of
three basis states $|\sigma\rangle_{j}$ with
$S_j^{z}|\sigma\rangle_{j}=\sigma|\sigma\rangle_{j}$,
$\sigma=0,\pm1$. It is convenient to write down the spin-1 state at
site $j$ as
\begin{equation}
\label{states}
|\psi\rangle_{j}=\sum_{a=x,y,z}
t_{j,a}|a\rangle_{j},
\end{equation}
using the ``cartesian'' states
$|z\rangle=|0\rangle$,
$|x\rangle=(|-1\rangle-|+1\rangle)/\sqrt{2}$,
$|y\rangle=i(|-1\rangle+|+1\rangle)/\sqrt{2}$, then the three numbers
$t_{ja}$ transform under rotations as the components of a complex vector
$\vec{t}_{j}$.
The normalization condition brings the constraint
$\vec{t}_{j}^{*}\cdot\vec{t}_{j}=1$.
The states (\ref{states}) can be viewed as $SU(3)$ coherent states corresponding
to the bosonic operators $\hat{t}_{j,a}$, and the $S=1$ operator can be
represented as
$S_{j}^{a}=-i\epsilon_{abc}\hat{t}^{\dag}_{j,b}\hat{t}^{\vphantom{\dag}}_{j,c}$.
Taking into account that the state (\ref{states})
may contain an arbitrary overall phase factor, one concludes that the order
parameter space  $\mathbb{M}$ of the problem is four-dimensional and isomorphic
to $CP^{2}$.

The lattice Lagrangian of the model expressed in terms of the complex unit
vector $\vec{t}$ takes the form
\begin{equation}
\label{lagr-cp2d}
\mathcal{L}=\sum_{j}i(\vec{t}_{j}^{*}\cdot
\partial_{t} \vec{t}_{j})-W,\quad W=\sum_{\langle ij\rangle}
\langle\widehat{h}_{i,j}\rangle,
\end{equation}
where the local Hamiltonian average $\langle\widehat{h}_{i,j}\rangle$ 
is given by
\begin{equation}
\label{t-energy}
\langle\widehat{h}_{i,j}\rangle = J (\vec{t}^{*}_{i}\cdot
\vec{t}^{\vphantom{*}}_{j})(\vec{t}^{*}_{j}\cdot \vec{t}^{\vphantom{*}}_{i})
+(J-K)(\vec{t}^{*}_{i}\cdot \vec{t}^{*}_{j}) (\vec{t}^{\vphantom{*}}_{i}\cdot
\vec{t}^{\vphantom{*}}_{j}).
\end{equation}
This makes obvious that the system is always invariant under global rotations
$t_{j,a}\mapsto \mathcal{R}_{ab}t_{j,b}$, with an arbitrary $O(3)$ rotation
matrix $\mathcal{R}$, as well as under local ``gauge'' transformation
$\vec{t}_{j}\mapsto \vec{t}_{j}e^{i\chi_{j}}$.  At $J=K$ the symmetry becomes
higher as there is an invariance under a global transformation
$\vec{t}_{j}\mapsto U\vec{t}_{j}$, with $U\in SU(3)$. Moreover, if the lattice
is bipartite, at $J=0$ the energy is invariant under making an arbitrary $SU(3)$
rotation on the sites belonging to one sublattice if this is accompanied by a
conjugate transformation $\vec{t}_{j}\mapsto U^{*} \vec{t}_{j}$ at the other
sublattice, so the point $J=0$ is $SU(3)$-invariant as well.

Breaking up the complex vector $\vec{t}=\vec{u}+i\vec{v}$ into two real
vectors representing its real and imaginary parts, one can write the on-site spin
and quadrupole averages as
\begin{eqnarray}
\label{aver}
&& \langle \vec{S}\rangle=2(\vec{u}\times \vec{v}),\nonumber\\
&& S_{ab}\equiv\langle S_{a}S_{b}+S_{b}S_{a}\rangle=2(\delta_{ab}-u_{a}u_{b}-v_{a}v_{b}).
\end{eqnarray}

One can use a different parametrization, directly connected to the physical
averages, by introducing the eight-component vector $\vec{n}$,
\begin{equation}
\label{8-vector}
n_{\alpha}=\mbox{tr} (\vec{t}^{*}\cdot
\widehat{\lambda}_{\alpha} \vec{t}),
\end{equation}
where $\widehat{\lambda}_{\alpha}$, $\alpha=1,\ldots 8$ are the well-known
Gell-Mann matrices which form, together with a unit matrix $\openone$, a basis
in the $SU(3)$ matrix space. The vector $\vec{n}$ is subject to the following
two constraints:
\begin{equation}
\label{8-constr}
\vec{n}^{2}=4/3,\quad \vec{n}\cdot (\vec{n} * \vec{n})=8/(3\sqrt{3}),
\end{equation}
where the $*$-product  of any two vectors $\vec{n}$ and
$\vec{n}'$ is defined as $(\vec{n} *
\vec{n}')_{\alpha}=\sqrt{3}d_{\alpha\beta\gamma}n_{\beta}n'_{\gamma}$,
and $d_{\alpha\beta\gamma}$ are the structure constants
defined by the anticommutation properties of the Gell-Mann matrices,
$\{ \lambda_{\alpha},\lambda_{\beta}
\}=\frac{4}{3}\delta_{\alpha\beta} \openone
+2d_{\alpha\beta\gamma}\lambda_{\gamma}$. One can show that the
constraints (\ref{8-constr}) in fact reduce the dimension of the
$\vec{n}$-space to four. The quantities $n_{\alpha}$ correspond to
the following on-site averages:
\begin{eqnarray}
\label{8-aver}
&&n_{2}=\langle S_{z}\rangle, \quad
 n_{5}=-\langle S_{y}\rangle,\quad
 n_{7}=\langle S_{x}\rangle, \nonumber\\
&& n_{4}=S_{xz}, \quad
 n_{6}=S_{yz},\quad n_{1}=S_{xy} \\
&& n_{3}=(S_{xx}-S_{yy})/2,\quad 
n_{8}=\sqrt{3}\big(S_{zz}/2 -2/3\big),\nonumber
 \end{eqnarray}
which can be split into the vector of
spin averages $\vec{m}$
and the vector of quadrupolar averages $\vec{d}$,
\begin{equation}
\label{qm}
 \vec{m}=(n_{7},-n_{5},n_{2}),\quad
\vec{d}= (n_{1},n_{3},n_{4},n_{6},n_{8}).
\end{equation}
In those variables, the Hamiltonian takes the simple form
\begin{equation}
\label{8-energy}
\langle h_{i,j}\rangle=-\frac{K}{3}-\frac{K}{2} (\vec{d}_{i}\cdot\vec{d}_{j})
+\frac{1}{2}(K-2J)(\vec{m}_{i}\cdot\vec{m}_{j}),
\end{equation}
which explicitly shows that $J>K$ corresponds to a ferromagnet (FM),
$J<0$ to an antiferromagnet (AFM), and $0<J<K$ to a quadrupolar
(spin nematic) order (hereafter we assume that $K>0$ and will not
discuss the so-called orthogonal spin nematic present at $K<0$).

In terms of $\vec{n}$, the lattice Lagrangian can be written as
$\mathcal{L}=\sum_{j}\Phi(\vec{n}_{j}) -\sum_{\langle ij\rangle} \langle
h_{ij}\rangle$, with the dynamic part
\begin{equation}
\label{8-dyn}
\Phi(\vec{n})=\frac{3}{4}\frac{\vec{n}_{0}\cdot (\vec{n}\wedge
  \partial_{t}\vec{n})}{1+\frac{3}{2}\vec{n}_{0}\cdot\vec{n}}.
\end{equation}
Here the $SU(3)$-crossproduct is defined as
$(\vec{n}\wedge\vec{n}')_{\alpha}=f_{\alpha\beta\gamma}n_{\beta}n'_{\gamma}$,
where $f_{\alpha\beta\gamma}$ is another set of structure constants defined
by commutators of the group generators
$[\lambda_{\alpha},\lambda_{\beta}]=2if_{\alpha\beta\gamma}\lambda_{\gamma}$,
and $\vec{n}_{0}$ is an arbitrary vector satisfying the constraints
(\ref{8-constr}).

\emph{Topological analysis.--}  To describe topological
solitons, one needs to pass to the continuum description first. The
continuum Lagrangian of the model \eqref{lagr-cp2d}  can be obtained by the gradient expansion
of the discrete energy $W$ retaining the leading terms, that gives $W=\int
d^{2}x\, w$ with
\begin{multline}
\label{lagr-cp2} w=J\{ |\partial_{\mu}\vec{t}|^{2} -
|\vec{t}^{*}\cdot\partial_{\mu}\vec{t}|^{2}\} +  (J-K)|\vec{t}^{2}|^{2} \\
- (J-K)\Big\{ |\vec{t}\cdot
\partial_{\mu}\vec{t}|^{2}
+\frac{1}{2}\big[\vec{t}^{2}(\partial_{\mu}\vec{t})^{2}+
\mbox{c.c}\big]\Big\},
\end{multline}
where $\mu$ runs over space coordinates $(x,y)$. The above form is valid for the  region $J>K/2$,
where the short-range spin-spin correlations are of the ferromagnetic
type, as can be seen from (\ref{8-energy}).

To classify the topological excitations, one needs to know the
so-called degeneracy space $\mathbb{M}_{D}$ that includes all
values of the order parameter field corresponding to the ground
state of the system. For the model (\ref{ham}) the space
$\mathbb{M}_{D}$ is continuous and depends on the type of the ground
state: for FM or AFM it coincides with the unit sphere $S^{2}$, for
the nematic case it is a 2d real projective space
$RP^{2}=S^{2}/Z^{2}$ (a unit sphere with the opposite points identified), and at $J=K$ the
degeneracy space is enlarged to $CP^{2}$.  For all the above spaces,
the second homotopy group is nontrivial,
$\pi_{2}(\mathbb{M}_{D})=\mathbb{Z}$ which makes possible 
the existence of so-called localized topological solitons,
whose order parameter distribution becomes uniform away
from some point. 

If the order parameter lies completely in $\mathbb{M}_{D}$, the energy contains
only terms with gradients, so there is no natural space scale. If corresponding
soliton solutions exist, they have a finite energy which does not depend on
their size, and are stable against collapse.  Another possibility is to allow
the order parameter to leave $\mathbb{M}_{D}$, which breaks the scale
invariance.  Static solitons of that type are unstable against
collapse due to  the Hobart-Derrick theorem, but they can be
stabilized by some internal dynamics \cite{manton,Kosevich+All}.  We
will study the structure of both types of solitons for the model (\ref{ham}).

For the sake of analyzing static soliton solutions the Lagrangian
(\ref{lagr-cp2d}) with the energy \eqref{lagr-cp2} is equivalent to
the 2d $CP^{2}$ model \cite{DAdda+78} with an additional
``anisotropy term'' proportional to
$(J-K)$. Let us start from the $SU(3)$-symmetric point $J=K$. In
that case a localized topological soliton corresponds to the field
configuration with nonzero topological charge \cite{DAdda+78}:
\begin{equation}
\label{charge-su3}
q=-\frac{i}{2\pi}\int d^{2}x
\epsilon_{\mu\nu}(\partial_{\mu}\vec{t}^{*}\cdot\partial_{\nu}\vec{t}),
\end{equation}
where the indices $\mu$, $\nu$ run over $(x, y)$.  The invariant
(\ref{charge-su3}) takes only integer values and corresponds to the mapping
of the compactified 2d space $S^{2}$ onto $CP^{2}$. The exact
$q=1$ soliton solution is well known \cite{DAdda+78}:
\begin{equation}
\label{q1sol}
\vec{t}=(\xi\vec{a}+z\vec{b})/\sqrt{|z|^{2}+\xi^{2}},
\end{equation}
where $z=x+iy$ is the complex coordinate (the soliton center is assumed to be at
the origin), $\vec{a}$ and $\vec{b}$
are two mutually orthonormal complex vectors, and $\xi$ has the
meaning of the soliton size. The energy of such excitation according to
(\ref{lagr-cp2}) is $E=2\pi K$. For an arbitrary value of $q$, the
general soliton solution can be written as
\begin{equation}
\label{gen-sol}
t_{a}=\frac{f_{a}}{(\sum_{a}|f_{a}|^{2})^{1/2}},\quad
f_{a}=c_{a}\prod_{k=1}^{q}(z-z_{k,a}),\quad a=x,y,z,
\end{equation}
and the corresponding energy is $E=2\pi K |q|$.

\emph{Ferromagnetic solitons.--} On the ferromagnetic side $J>K$ the minimum of  energy is
achieved for
\begin{equation}
\label{t-FM}
\vec{t}=(\vec{e}_{1}+i\vec{e}_{2})/{\sqrt{2}}
\end{equation}
with $\vec{e}_{1,2}$ being a pair of orthogonal real unit vectors.  In that case
on the degeneracy space $\mathbb{M}_{D}$ the order parameter is equivalent to
the unit vector $\vec{m}=(\vec{e}_{1}\times \vec{e}_{2})$ (a rotation 
around $\vec{m}$ corresponds
to a change of the overall phase factor $\vec{t}\mapsto
\vec{t}e^{i\varphi}$ and thus does not change the physical state).  Thus,
localized topological solitons for $J>K$ correspond to the mapping $S^{2}\mapsto
S^{2}$ and are characterized by another topological charge
\begin{equation}
\label{charge-o3}
Q_{m}= \frac{1}{8\pi}\int d^{2}x\,
\varepsilon_{\mu\nu} \vec{m}\cdot(\partial_{\mu}\vec{m}\times \partial_{\nu}\vec{m}).
\end{equation}
It is easy to calculate the topological charge
(\ref{charge-su3}) for a  restricted field configuration satisfying
(\ref{t-FM}): a general pair of orthonormal vectors $\vec{e}_{1,2}$ can be
obtained from $\vec{e}_{x,y}$ by an arbitrary rotation
$\mathcal{R}(\theta,\varphi,\psi)$, where 
$\theta$ and $\varphi$ are respectively the polar and azimuthal angles
characterizing the direction of the unit magnetization vector $\vec{m}$, and the
third angle $\psi$ corresponds to the rotation around $\vec{m}$. A
straightforward calculation yields
\begin{equation}
\label{q=2Q}
q=\frac{1}{2\pi}\int d^{2}x\sin\theta
\epsilon_{\mu\nu}(\partial_{\mu}\theta)( \partial_{\nu}\varphi)=2Q_{m}.
\end{equation}
One is led to conclude that solitons of the $CP^{2}$ model tend to
pair upon perturbing the $SU(3)$ symmetry, which constitutes the central
observation of the present paper.

The above result can be also obtained  by noticing
that for the configurations (\ref{t-FM}) the energy takes the form
$W=(J/2)\int d^{2}x\,(\partial_{\mu}\vec{m})^{2}$.
This is exactly  the energy of the $O(3)$ NLSM, and the well-known Belavin-Polyakov (BP) soliton solution
\cite{BelavinPolyakov75} with the
topological charge $Q_{m}=1$ will have the energy $E= 4\pi J$, which in the limit
$J\to K$ is twice the energy of the $q=1$ soliton (\ref{q1sol}) of the $CP^{2}$
model. In fact, one can explicitly check that the ferromagnetic BP soliton is a
particular case of the general  solution (\ref{gen-sol}) with $q=2$.

\emph{Solitons for spin nematic.--} On the nematic side $J<K$ the
minimum of energy is reached for $\vec{t}=\vec{u}e^{i\chi}$, where
$\vec{u}$ is a real unit vector and $\chi$ is an arbitrary phase. The degeneracy space 
is thus $\mathbb{M}_{D}=RP^{2}$. 
The energy then takes the form
$W=K\int d^{2}x\, (\partial_{\mu}\vec{u})^{2}$,
where $\vec{u}$ must be understood as a director, i.e., $\vec{u}$
and $-\vec{u}$ are physically identical. 
It is worth
noting that in contrast to the other phases the spin nematic
allows for a nontrivial $\pi_1$-topological charge as well, $\pi_{1}
(RP^{2})=Z_2$.

If one defines the topological charge $Q_{u}$
according to (\ref{charge-o3}), simply replacing $\vec{m}$ by
$\vec{u}$, then in the BP soliton with $Q_{u}=1$ the
director $\vec{u}$ goes over $\mathbb{M}_{D}$ twice; the energy of
such a solution is $E_{BP}=8\pi J$. However, the director property
of $\vec{u}$ allows one to construct a solution
\cite{IvanovKolezhuk03} with $\vec{u}$ going over $\mathbb{M}_{D}$
just once, which has $Q_{u}=\frac{1}{2}$ and the energy
$\widetilde{E}_{BP}=4\pi J$. In the limit $J\to K$ this is again
twice as much as the energy of the $q=1$ solution (\ref{q1sol}), 
which suggests that this soliton is a
descendant of the $q=2$ solution of the $CP^{2}$ model.  This
indicates that the tendency to pairing exists on the nematic
side as well.

\emph{The fate of solitons with \ $q=1$.--} 
Up to now we have considered only static solitons with the order parameter lying
completely inside $\mathbb{M}_{D}$. We found that for  $J-K\not=0$ the lowest energy 
solutions of that type are  descendants of  $q=2$ soliton of the $CP^{2}$ model,
while the $q=1$ solution seems to exist only at $J=K$. 
To get further understanding of what
happens in the vicinity  of the $SU(3)$-symmetric point $J=K$, let us discuss
the $CP^2$-soliton with $q=1$ for small but finite $J-K$.
One can easily see that at $J-K \neq 0$ any solutions with $q=1$ must involve
a deviation of the order parameter from the degeneracy space $\mathbb{M}_{D}$. 
Due to the Hobbart-Derrick
theorem, this means instability of static solitons with $q=1$
against collapse.  However, $q=1$ solitons can be
stabilized by internal dynamics in presence of additional
integrals of motion, e.g.,  stable solitons with
the magnetization vector precessing around the easy axis exist in the uniaxial ferromagnet
\cite{Kosevich+All}.
In our case, it is also possible to construct
such a solution. In terms of the complex
vector $\vec{t}=\vec{u}+i\vec{v}$ this is a planar configuration,
where $\vec{u}$ and $\vec{v}$ are parallel to the plane $(1,2)$
orthogonal to some axis $\vec{e}_{3}$, for definiteness let it be
the $z$ axis (a more general solution can be
obtained by an arbitrary rotation). It is convenient to use the $8$-vector
notation  (\ref{8-vector}): only four  components of $\vec{n}$ are
nonzero and it takes the form
\begin{equation}
\label{R-ansatz} \vec{n}=\big(R_{x},
R_{z}, R_{y},
0,0,0,0,1/\sqrt{3}\big),
\end{equation}
where $\vec{R}$ is a unit vector combining one spin average
$R_{z}=m_{3}$ and two quadrupolar variables $R_{x}=d_{1}$,
$R_{y}=d_{2}$ (cf.\ (\ref{qm})). 
Using (\ref{8-energy}) and (\ref{8-dyn}), one obtains the effective
Lagrangian for the chosen subspace,
\begin{eqnarray}
\label{R-lagr} \mathcal{L}_{R}&=&\frac{1}{2}\sum_j
\frac{\vec{R}_{0}\cdot(\vec{R}_j\times
  \partial_{t}\vec{R}_j)}{1+\vec{R}_{0}\cdot\vec{R}_j} -W_{R}\\
W_{R}&=&=-\sum _{<ij>} \bigl[\frac{K}{2}\vec{R}_i \vec{R}_j +
(J-K)R_{z,i}R_{z,j}\bigr]\, .\nonumber
\end{eqnarray}
where $\vec{R}_{0}=(0,0,-1)$, and in (\ref{8-dyn}) we have used
$\vec{n}_{0}=(0,0,-1,0,0,0,0,\frac{1}{\sqrt{3}})$. The
Lagrangian (\ref{R-lagr}) describes the dynamics of a classical
anisotropic ferromagnet with the unit magnetization vector
$\vec{R}$; the anisotropy constant is proportional to $J-K$.  At the
isotropic point $J=K$ the energy $W_{R}=(K/2)\int (\nabla
\vec{R})^{2}d^2x$, and there exists a BP-type soliton  that has the energy
$E_{J=K}=2\pi K$ and is a special case of the $q=1$ $CP^{2}$
solution (\ref{q1sol}).
The $CP^{2}$ charge $q$ given by (\ref{charge-su3}) is obviously
equal to the Pontryagin index $Q_{R}$ defined by (\ref{charge-o3}) with
$\vec{m}\mapsto\vec{R}$; the BP solution corresponds to the
mapping of $S^{2}$ onto the subspace
$CP^{1}$ embedded into $CP^{2}$ and has $Q_{R}=q=1$.

\begin{figure}[tb]
\includegraphics[width=0.4\textwidth]{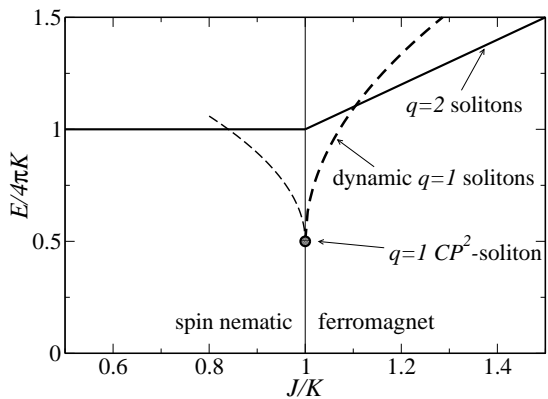}
\caption{\label{fig:energy} Energy of topological solitons in the
vicinity of the $SU(3)$-symmetric point $J=K$. Solid lines show the
energy of static solitons with the $CP^{2}$ topological charge $q=2$, and
the dashed line corresponds to the descendants of the $q=1$ soliton
of the $CP^{2}$ model.}
\end{figure}

For a finite ``anisotropy'' $(J-K)$ the BP soliton becomes unstable against
collapse, but the situation is different for the spin-nematic and FM regions.
In the FM case ($J>K$) the anisotropy is of the easy-axis type, and there exist
$Q_{R}=q=1$ dynamic solutions, with $\vec{R}$ precessing around the $z$ axis
\cite{Kosevich+All}, which are smoothly connected to the BP solitons in the
$J\to K$ limit.  A detailed analysis \cite{IvMerkSZasp} shows that the minimal
energy of such dynamic solitons exhibits a nonanalytical behavior of the type
$E_{\rm min}=E_{J=K}(1+3.74\sqrt{J/K-1})$, as shown in Fig.\ \ref{fig:energy}.
For small $J-K \leq 0.1K$, when the above expression is valid, the
energy of a static $q=2$ ($Q_{m}=1$) soliton considered above  stays
higher than the energy of the $q=1$ ($Q_{R}=1$) dynamical soliton, but at the
same time, it remains smaller that the energy of two dynamical $q=1$ solitons,
which indicates that it is energetically favorable to bind two $Q_{R}=1$
solitons into a single $Q_{m}=1$ one.

In the nematic case ($J<K$) we effectively have a ferromagnet with the
easy-plane anisotropy. For such case, delocalized $\pi_1$-solitons (vortices)
exist. Vortices in $\vec{R}$-field correspond to spin-nematic disclinations
considered in Ref.\ \cite{IvanovPZh06}.  The energy of a single vortex diverges
logarithmically with the system size, so a static vortex-antivortex pair is
unstable against collapse.  The BP soliton can be considered as a pair of
``merons'' carrying topological charge $Q_{R}=\frac{1}{2}$ each
\cite{Affleck89rev}.  For small $(K-J)$ those ``merons'' can be viewed as a
vortex and antivortex with a finite out-of-plane component of the vector
$\vec{R}$, they are subject to a gyroforce \cite{IvanovPZh06}, and there may
exist stable dynamic solutions (rotational pairs of vortices) similar to those
studied in Ref.\ \cite{Gary+pary,Shoorik}.  Their energy will tend to $2\pi K$
in the limit $J\to K$; similarly to the FM case, in the vicinity of the $J=K$
point the $Q_{R}=1$ topological solitons will be unstable against pairing into
``nematic'' Belavin-Polyakov solitons with $Q_{u}=\frac{1}{2}$.

Finally, a few words are to be said about the other,
antiferromagnetic $SU(3)$-symmetric point $J=0$.  From
(\ref{t-energy}) one can see that on any bipartite lattice the
transformation $\vec{t}_{j}\mapsto \vec{t}^{*}_{j}$ for all $j$
belonging to one sublattice maps the points $J=0$ and $J=K$ onto
each other.  As can be seen from (\ref{8-energy}), and is especially
clear from the ``spin analogy'' \eqref{R-lagr}, for $J<K/2$ the
short-range correlations are antiferromagnetic, and the proper
transition to the continuum description becomes more complicated;
however, one can show that the difference concerns only dynamics and
does not affect the static properties. The arguments leading to
(\ref{q=2Q}) and thus the conclusion on soliton pairing equally
apply to the vicinity of $J=0$ point. 

\emph{Summary.--} We have studied the structure of topologically nontrivial
solitons in a general non-Heisenberg model of the 2d isotropic $S=1$
magnet. In the vicinity of special points with $SU(3)$ symmetry 
the system can be described with the help of the
$CP^{2}$ model.  It is shown that when the
$SU(3)$ symmetry is broken down to $SU(2)$, solitons of the $CP^{2}$ model with
odd topological charge become unstable and bind into pairs.

\emph{Acknowledgments.--} AK was supported by the Heisenberg Program
Grant No.\ KO~2335/1-2 from Deutsche Forschungsgemeinschaft. BI and
RK were supported in part by the grant INTAS-05-1000008-8112 and by the
joint grant Ô25.2/081 from the Ministry of Education and Science of
Ukraine and Ukrainian State Foundation of Fundamental Research.

\end{document}